\documentclass[structabstract]{aa}  

\usepackage{natbib}
\bibpunct{(}{)}{;}{a}{}{,}
\usepackage{graphicx}
\usepackage{txfonts}
\usepackage{multirow} 
\usepackage{subfig}
\usepackage{url} 

\def\msun{$M_{\odot}$}
\def\sette{XTE~J1751--305}
\def\otto{XTE~J1814--338}
\def\nove{XTE~J0929--314}
\def\sax{SAX~J1808.4--3658}

\def\farcs{\hbox{$.\!\!^{\prime\prime}$}}
\def\fsecs{\hbox{$.\mkern-4mu^{s}$}}

\def\gsim{\hbox{\raise0.5ex\hbox{$>\lower1.06ex\hbox{$\kern-0.94em{\sim}$}$}}}
\def\lsim{\hbox{\raise0.5ex\hbox{$<\lower1.06ex\hbox{$\kern-0.94em{\sim}$}$}}}

\begin{document}
   \title{Search for pulsations at high radio frequencies from accreting millisecond X-ray pulsars in quiescence}

   \author{M.N. Iacolina
          \inst{1,}\inst{2}
          \and
          M. Burgay\inst{2}
          \and
          L. Burderi\inst{1}
          \and
          A. Possenti\inst{2}
          \and
          T. Di Salvo\inst{3}
          }

   \institute{Universit\`a di Cagliari, Dipartimento di Fisica, 
              SP Monserrato-Sestu km 0.7, 09042 Monserrato (CA), Italy\\
              \email{iacolina@ca.astro.it}
         \and
             INAF-Osservatorio Astronomico di Cagliari,
             Loc. Poggio dei Pini, Strada 54, 09012 Capoterra (CA), Italy
         \and
             Universit\`a di Palermo, Dipartimento di Scienze Fisiche 
             ed Astronomiche, via Archirafi 36, 90123 Palermo, Italy\\
             }

   \date{Received 18 January 2010 / Accepted 31 May 2010}

  \abstract
   {It is commonly believed that millisecond radio pulsars have been spun up by transfer of matter and angular momentum from a low-mass companion during an X-ray active mass transfer phase. A subclass of low-mass X-ray binaries is that of the accreting millisecond X-ray pulsars, transient systems that show periods of X-ray quiescence during which radio emission could switch on.}
   {The aim of this work is to search for millisecond pulsations from three accreting millisecond X-ray pulsars, \sette, \otto, and \sax, observed during their quiescent X-ray phases at high radio frequencies (5 $\div$ 8 GHz) in order to overcome the problem of the free-free absorption due to the matter engulfing the system. A positive result would provide definite proof of the recycling model, providing
the direct link between the progenitors and their evolutionary products.}
   {The data analysis methodology has been chosen on the basis of the precise knowledge of orbital and spin parameters from X-ray observations. It is subdivided in three steps: we corrected the time series for the effects of (I) the dispersion due to interstellar medium and (II) of the orbital motions, and finally (III) folded modulo the spin period to increase the signal-to-noise ratio.}
   {No radio signal with spin and orbital characteristics matching those of the X-ray sources has been found in our search, down to very low flux density upper limits.}
   {We analysed several mechanisms that could have prevented the detection of the signal, concluding that the low luminosity of the sources and the geometric factor are the most likely reasons for this negative result.}


\titlerunning{Searching for pulsed radio emission from AMXPs}
\authorrunning{M.N. Iacolina et al.}
   \maketitle
%

\section{Introduction}

It is commonly believed that accreting millisecond X-ray pulsars (AMXPs), transient binary systems hosting a fast spinning 
($P_{\rm SPIN} \sim $ 1 ms) and weakly magnetised ($B \sim 10^{8 \div 9}$ Gauss) neutron star (NS), are the progenitors of the radio 
millisecond pulsars (MSPs), as argued by the recycling model \citep{acrs82,bv91}. This model asserts that the NS in these 
systems is spun up by the transfer of matter and angular momentum from its low-mass (M $\leq$ 1 M$_{\odot}$) companion star via the 
formation of an accretion disk. When this process ends, the NS switches on as a radio MSP.

A basic requirement for the switching on of the radio emission is that the space surrounding the NS has to be free of matter,
a condition that can be fulfilled during the quiescence phase of AMXPs. For this reason this phase constitutes the most promising one to 
investigate for confirming of the recycling model. The aim of this work is, in fact, to search for radio millisecond pulsations 
from a sample of AMXPs in their quiescence phase: a positive result would unambiguously establish that AMXPs are the progenitors 
of at least some of the radio MSPs.

In the past decade, from 1998 April, the date of the discovery of the first AMXP, \sax\ \citep{wv98}, several attempts have been made
to obtain this confirmation \citep[e.g.][]{bbp+03}, but, despite the eleven additional systems 
discovered since \citep[e.g.][]{gcmr02,cap+08}, we have not obtained a positive result yet.

A possible explanation of some of these failures has been given by \citet{bpd+01}, asserting that detection of radio pulsations 
from AMXPs could be hampered by matter surrounding the system. In fact, during the so-called {\it radio-ejection} phase, the pressure 
of the rotating magneto-dipole could prevent the infalling matter from the companion of the NS in the binary system from reaching the NS 
Roche lobe, forcing it to leave the system from the Lagrangian point $L_1$.
Once this happens, such matter, which is still carrying the angular momentum, will rotate around the two stars embedding the system.
Even if the pulsar radio emission was switched on and the system in X-ray quiescence, it could be absorbed by a free-free mechanism.
Since the optical depth for the free-free absorption, $\tau_{\rm ff}$, depends on the square inverse of the frequency, 
observations at higher frequencies could encompass this problem.

With this scenario in mind we have undertaken a campaign of observation at high radio frequencies (see Table \ref{tab:tauffall}) for 
four AMXPs, \nove, \sette, \otto, and \sax, in their quiescence phase.
Results for \nove\ were presented in \citet{ibb+09}. Here we used the same equation to estimate $\tau_{\rm ff}$ for the other three 
sources, and we obtained the values listed in Table \ref{tab:tauffall}. The parameters are: 
for the four sources, $m_1\sim 1.4$, $X = 0.7, Y = 0.3, \gamma \sim 1, T_4 \sim 1$;
for \otto, $\dot{m}_{-10}\sim 1.2$, $m_2\sim 0.17$, $P_{\rm h}=4.27$ \citep{ms03};
for \sette, $\dot{m}_{-10}\sim 8.5$, $m_2\sim 0.015$, $P_{\rm h}=0.71$ \citep{mss+02b}; 
for \nove, $\dot{m}_{-10}\sim 2.9$, $m_2\sim 0.02$, $P_{\rm h}=0.73$ \citep{gcmr02};
and for \sax, $\dot{m}_{-10}\sim 3.7$ (which is the average mass transfer rate for the 1998, 2000, and 2002 outbursts, see \citealt{pmb+05}), $m_2\sim 0.05$, $P_{\rm h}=2.01$.
In this table, the values of $\tau_{\rm ff}$ obtained at 1.4 
GHz (the typical frequency used to observe pulsar) are much higher than unity, so the radiation would be totally absorbed, 
while for frequencies higher than 5 GHz we obtained values smaller or close to unity for all the sources and, reasonably assuming 
that the matter enclosing the system is clumpy (i.e. there are favourable directions where the optical depth is lower than the 
average values of the Table), we have a higher probability of detecting the radio signal.
\begin{table}[ht]
\caption{Optical depth at various radio frequencies, for the four sources.}
\begin{center}
\begin{tabular}{lccc}
\hline

\hline

\hline
\hline
\multirow{2}{*}{Source} &\multicolumn{3}{c}{$\tau_{\rm ff}$}\\
\cline{2-4}
&1.4 GHz&6.5 GHz& 8.5 GHz\\
\hline
\otto&2.2&1.2&0.05\\
\sette&27.8&1.1&0.7\\
\nove&5&0.2&0.1\\
\hline

\hline
&1.4 GHz&5 GHz& 6 GHz\\
\hline
\sax&9.3&0.7&0.5\\
\hline
\hline
\end{tabular}
\end{center}
\label{tab:tauffall}
\end{table}

The unknown inclination of the systems negligibly affects the estimate of the optical depth, considering that, while the 
estimated companion mass only slightly increases for a decrease in the inclination, the amount of 
matter along the line of sight significantly decreases, because the disk is in the orbital plane.


\section{The sources, the observations, and the data analysis method}

Two series of radio observations were taken on 2003 December 20--23 for \sette\ and \otto, and on 2002 August 5--7 for
\sax, using the Parkes 64 m radio telescope. Observation parameters are listed in Table \ref{tab:parameters}. 
The data analysis methodology was chosen on the basis of the precise knowledge of the spin and orbital parameters from X-ray observation.
It is the same as was adopted for the source \nove\ presented by \citet{ibb+09}, where it is described in detail.

The original ephemerides were published by \citet{mss+02b} for \sette, by \citet{ms03} for \otto\ and by \citet{cm98} for \sax, 
and subsequently refined  by \citet{pdb+07,pmb+08} for \sette\ and \otto, and by \citet{hpc+09} (but also by \citealt{brd+09}, 
\citealt{dbr+08a} and \citealt{hpc+08}) for \sax. Table \ref{tab:parameters} reports the most updated ephemerides used in this work.

The first part of the data analysis was to correct the time series for the dispersion effects of the ISM; the steps, 
the ranges of DMs used, and the values of the local DMs for the three sources are indicated in Table \ref{tab:parameters}
(for \sax, we considered the highest mass transfer rate assumed in quiescence, $\dot{m} \sim 10^{-9}$ \msun/y, proposed by 
\citealt{dbr+08a}).
We then barycentred the data series to correct for the orbital effects considering the propagation of the uncertainties 
in the ephemerides derived from X-ray observations over the time range between X-ray and radio observations. 
This time range corresponds to $\sim$20000 orbits for \sette\ and $\sim$1000 orbits for \otto. For \sax, the adopted X-ray 
ephemerides are derived from the analysis of the secular evolution reported by \citet{hpc+09} (see the Table \ref{tab:parameters}), 
which refers to the time of $\sim$100 orbits after the radio observations.

For \sette, only the propagation of the orbital period error (90\% confidence level) affected the time series losing the possibility of 
detecting the signal: i.e. producing a broadening of the pulse of 0.9 in pulsar phase. To reduce this broadening to at 
most 0.1 in pulsar phase, one has to search for the signal at 18 trial values of the orbital period, 9 above and 9 below the nominal 
value, covering the uncertainty range.

For \otto\ and \sax, the uncertainty in all the parameters within the 90\% confidence level for \otto\ and 1$\sigma$ level for \sax\ 
does not affect the detectability of the pulsation. We then corrected the data series by only using the nominal values of the parameters.

The last step in this search is to fold the time series using the spin parameters reported in Table \ref{tab:parameters}. 
The spin period range explored has to be chosen by considering the nominal value of the spin period at the epoch of the radio
observations, as explained by \citet{ibb+09}.

For \sax, which is the only AMXP that showed more than one outburst, we considered the value of the spin period at the epoch of the 
X-ray observations, $P_{\rm X}$, derived by \citet{hpc+09} and the value of its derivative, $\dot{P}_{\rm X}$, measured during the outburst 
closest to the radio observations time, reported by \citet{bdm+06} and listed in Table \ref{tab:parameters}, which turns out to be
higher than what is derived by the analysis of the secular evolution obtained by \citet{hpc+09} and constitutes an upper limit in the 
search (see below). The folding trial values are indicated in Table \ref{tab:parameters}.

To check the plausibility of the adopted spin period interval, for \otto, we derived the spin period 
derivative, $\dot{P}_{\rm dip}$, through an estimate of the surface magnetic field, considering the magnetic torque acting on the neutron 
star, as discussed by \citet{ibb+09}. The value obtained is $\dot{P}_{\rm dip} = 6 \times 10^{-20}$, which is lower than $\dot{P}_{\rm X}$ 
and then safely contained in our interval of spin period trial values.
For \sax, \citet{bdm+06} calculated a value of the magnetic field equal to $B_{\rm S}\sim (3.5 \pm 0.5) \times 10^8$ Gauss. 
The spin period derivative, therefore, results $\dot{P}_{\rm dip} = 5.5 \times 10^{-20}$, which is once again $< \dot{P}_{\rm X}$. 

For \sette\ \citet{whh+05} derived an estimate of the magnetic field during its quiescence phase constrained to be
$ < (2.5 - 6) \times 10^8$ G, using the value reported in Table \ref{tab:parameters} for the distance.
The spin period derivative resulted to be $ < (2.7 - 15) \times 10^{-20}$ which is lower than 
$\dot{P}_{\rm X}$, consequently, the adopted interval of spin periods is again safely large.


\section{Results}
At the end of the three steps of analysis we produced $\sim$67000 folded profiles for \sette, $\sim$330 for \otto, and $\sim$200 for \sax,
reporting the results from the folding of the dedispersed, deorbited, and barycentred time series.
The best ones were displayed for visual inspection to search for possible pulsar suspects.

\begin{figure}[tt]
\begin{center}
\includegraphics[scale=0.33]{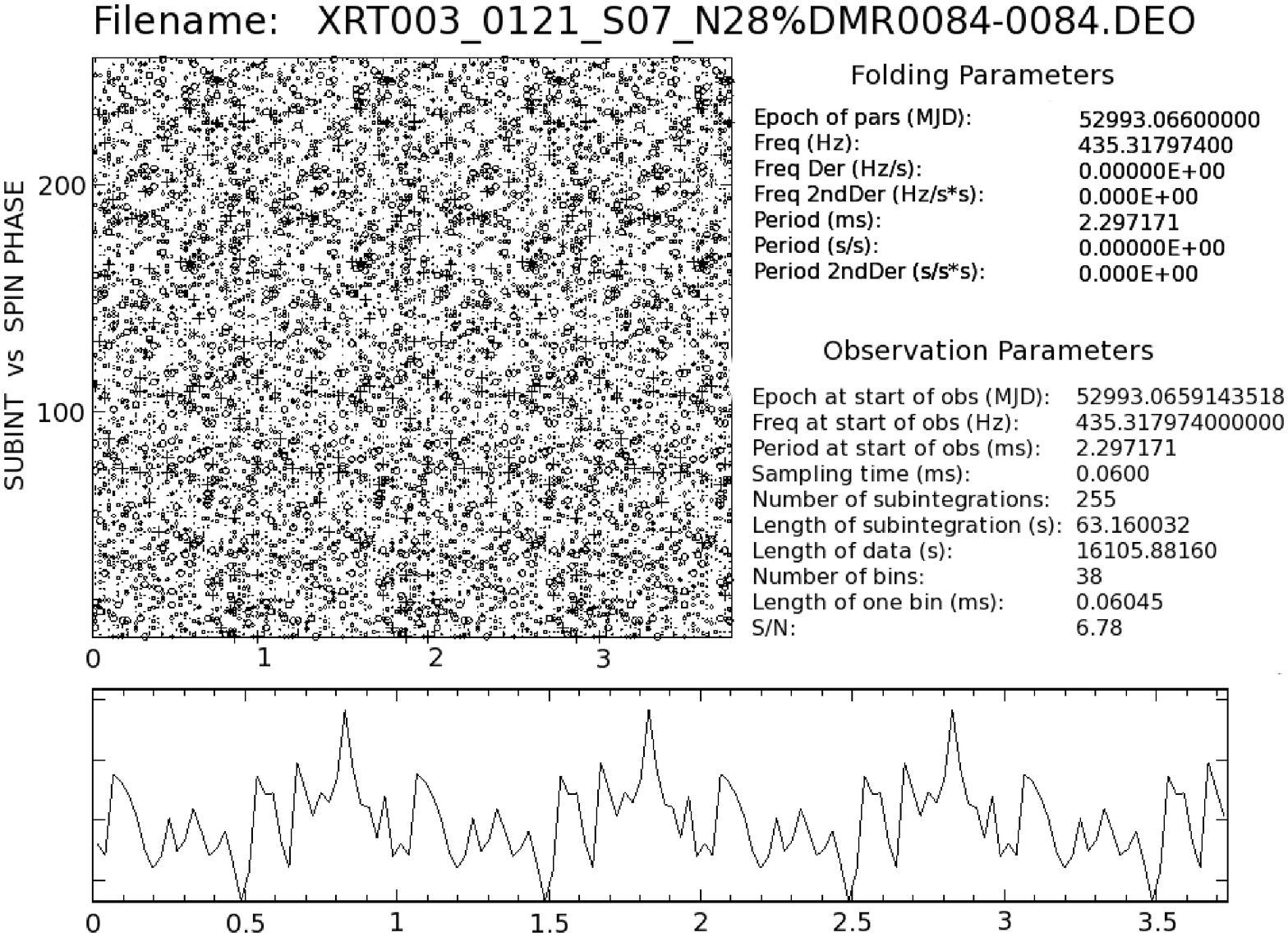}\\
\includegraphics[scale=0.44]{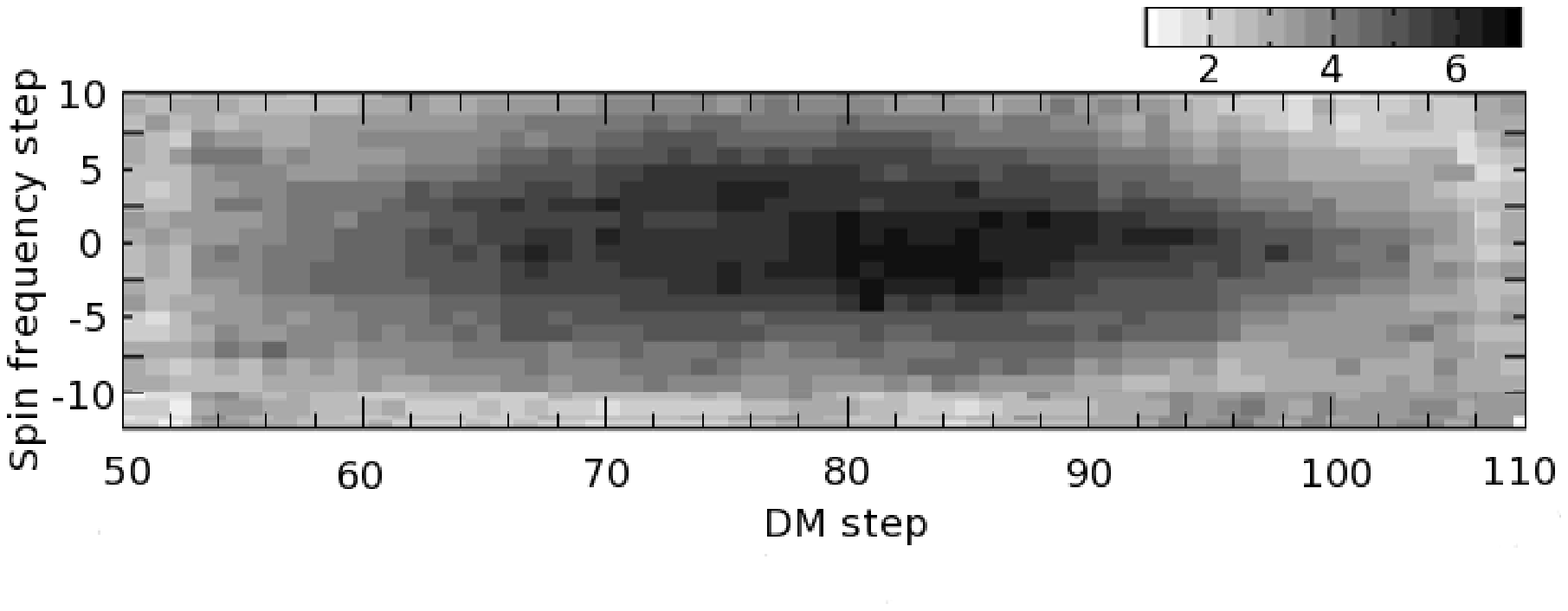}\\
\end{center}
\caption{{\it Upper panel}: Plot with the highest $S/N$ for \sette. {\it Lower panel}: $S/N$ in function of spin frequency and DM.
See the text for further explanations.}
\label{fig:best1751}
\end{figure}

For \sette, the highest $S/N$ obtained was 6.78 corresponding to a peak at 4$\sigma$ significance, the corresponding plot is shown in 
Fig. \ref{fig:best1751} (upper panel), where the grayscale on the left shows the signal in the 255 subintegrations in which the whole 
observation was split vs the spin phase, while the parameters used for the folding are indicated on the right. 
The diagram at the bottom displays about 4 phases of the integrated pulse profile. The parameters for the pulsar suspect are 
$DM$ = 735.41 pc cm$^{-3}$, $P_{\rm orb}$ = 2545.34361 s, $\nu_{\rm S}$ =  435.317974 Hz. This peak has a 40\% probability
of not being randomly generated over the 40755 trial foldings of the dataset corresponding to one of the two observations at 8.5 GHz.
Analysing the behaviour of the $S/N$ as a function of the spin frequency and DM trial values, a decreasing trend from a peak at $S/N =$ 
6.78 is found. Since this peak was at the limit of our search interval in spin frequency, we considered it appropriate to investigate 
for 11 more steps in the spin frequency, discovering, in this way, the whole trend: the maximum, which
corresponded to the same $S/N$, is defined in all the directions. The result is displayed in Fig. \ref{fig:best1751} (lower panel), 
where in the y-axis we have the spin frequency (10 steps above and 11 below the nominal value, corresponding to 0) and in the x-axis,
the DM (60 steps corresponding to the interval between 485.22 and 932.54 pc cm$^{-3}$) plotted for an orbital period 
$P_{\rm orb}=2545.34361$ s. A clear maximum is well defined, hence supporting the plausibility of the suspect.

This result is not confirmed in the other observation (at the same frequency) elaborated with the same parameters, but this could be due
to the clumpiness of the matter around the system. This suspect may thus deserve additional investigation in the future.

The highest $S/N$s reached for \otto\ and for \sax\ were 4.67 and 4.41, corresponding to a peak at 2.6$\sigma$  and 2.2$\sigma$ 
significance, respectively, with a probability of being real radio signals and not produced by noise of 18\% and 20\% over the 186 and 
57 trial foldings of their single dataset, respectively.
Unfortunately, visual inspection of these two results and others at lower $S/N$s did not provide any evidence of the pulsed signal, and
no other investigations have shown the positive signs in the trend of the $S/N$ obtained for \sette.
Finally, the observations at the other frequency, folded with the same parameters, did not displayed any suitable shape for the signal.
For these two sources, we can thus conclude that no pulsed radio signal has been detected at their spin period in our observations.

\subsection{Upper limit on the flux density}

Considering the parameters indicated in Table \ref{tab:parameters} and a duty cycle, $W/P_{\rm S}$, of 15\%, we estimated the upper limit 
on the flux density for the three observed sources at the nominal DM with the Eq. 9 in \citet{ibb+09}.
Including results for \nove\ \citep{ibb+09}, we obtained the values indicated in Table \ref{tab:sensibility}.

\begin{table}[ht]
\caption{Flux density upper limits for the four sources at the analysed frequencies.}
\begin{center}
\begin{tabular}{lcc}
\hline

\hline

\hline
\hline
\multirow{2}{*}{Source} &\multicolumn{2}{c}{$S_{\rm max}$ ($\mu$Jy)}\\
\cline{2-3}
&6.5 GHz& 8.5 GHz\\
\hline
\otto&52&25\\
\sette&--&31 - 30\\
\nove&68&26\\
\hline
&5 GHz& 6 GHz\\
\cline{2-3}
\sax &59&59\\
\hline
\hline

\hline

\hline
\end{tabular}
\tablefoot{For \sette\ the two observations are both at 8.5 GHz.}
\end{center}
\label{tab:sensibility}
\end{table}

\section{Discussion}

In this section we discuss the results of our findings, including results for \nove\ presented in \citet{ibb+09}.
A part for the result obtained for the 8.5 GHz observation of \sette, deserving additional investigation, no radio pulsed emission has 
been found above the reported upper limits in the data analysed. Assuming that the radio emission was switched on during the observations 
of the three sources, we investigated the possible reasons that prevented detecting of the radio signals.

Considering Eq. 11 in \citet{ibb+09}, we estimated the probability that the beams of the sources do not intersect
our line of sight. Assuming a typical duty cycle of 15\% for each source, the probability that the beams of all the sources (including 
\nove) are missing our line of sight is $\sim$19\%. In order to exclude this geometric bias we would have to analyse the 
whole known sample of twelve AMXPs, and the probability of missing all the beams of the whole sample will drop to about 0.1\%.

\begin{figure}[tt]
\begin{center}
\includegraphics[angle=-90,scale=0.34]{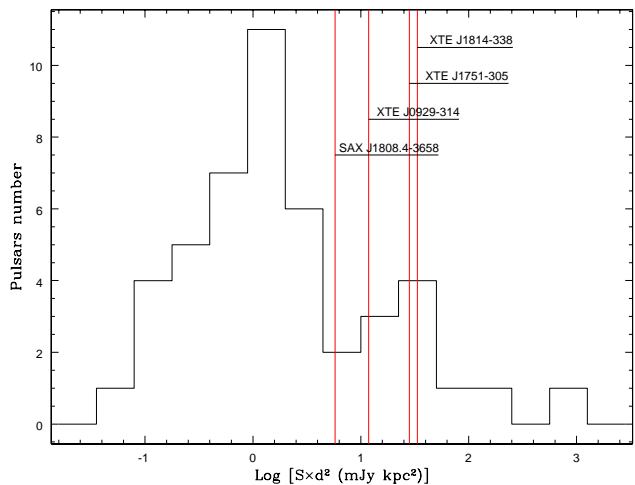}
\end{center}
\caption{Pseudoluminosity distribution of a sample of 46 MSPs in the galactic field. The red lines indicate the upper limits related to the four analysed sources.}
\label{fig:pseudolumi}
\end{figure}

The second possible reason that might have prevented the detection of a radio signal could be the low luminosity of the sources.
As the true luminosity of pulsars cannot be measured because of the unknown beaming fraction, a 'pseudo luminosity', $L$, is defined as 
the observed flux density, $S$, multiplied by the square of the pulsar distance, $d$ \citep{tm77}: $L=S \times d^2$.

In Fig. \ref{fig:pseudolumi} the logarithmic distribution of the pseudoluminosity at 1.4 GHz of the sample of 46 known galactic field MSPs is shown\footnote{Data taken from the ATNF pulsar catalogue -- \protect\url{http://www.atnf.csiro.au/research/pulsar/psrcat/}; \citet{mht+05}.}. The vertical red lines indicate the lower values of the upper limits on $L$ of \sette, \otto, \sax, and \nove\ (for the observations at 8.5 GHz and 5 GHz), scaled at 1.4 GHz, considering a dependence on the frequency $S(\nu) \propto \nu^{-\alpha}$, with an index $\alpha = 1.7$, and for the distances indicated in Table \ref{tab:parameters}.

These limits determine the probability that the true pseudo luminosity of each source is too faint for detection in our search. We calculated that this probability is about 90\% for \otto\ and \sette\ and $\sim$80\% for \sax\ and for \nove. The combined probability of the non 
detection due to the luminosity threshold of our survey is $\sim$50\%.
This percentage is not negligible and can be reduced by a deeper search and/or by a larger sample. 

Expanding the sample to all twelve known AMXPs, with a probability equal to about 80\% for each one, the combined probability would 
be $\sim$10\%, not enough for a safe detection. For a combined probability less than $\sim$0.1\%, 
we also have to perform a deeper search. The upper limit in pseudo-luminosity for each source having such a probability is 
$L \approx 2$ mJy kpc$^2$. The limit in flux for an average distance of 7 kpc becomes $S \approx  0.04$ mJy at 1.4 GHz which, 
scaled at 4.7 GHz (as, for example, for the observation of \sax), becomes $S \approx  0.003$ mJy. 
Such a limit can be reached by performing observations using telescopes with a larger bandwidth and higher instantaneous sensitivity.
In fact, a 4.7 GHz observation of \sax, performed using a 2 GHz bandwidth and 2.01 K Jy$^{-1}$ gain obtainable at the 
Green Bank radio telescope (GBT), would have reached a flux density limit of $\approx$0.003 mJy.


\section{Conclusions}
We performed a search for millisecond pulsations in three AMXPs, \otto, \sette, and \sax, in their quiescence phases at high radio
frequencies. Discussion was done by including the previous results obtained  by \citet{ibb+09} for \nove.
Except for the case of \sette, for which further investigations are needed, no pulsations with the expected 
periodicity have been detected in the analysed data. The flux density upper limits determined by our search (including \nove)
are indicated in Table \ref{tab:sensibility}.

Possible mechanisms that might have hampered the observation of the pulsed signal could concern the luminosity of the four analysed 
sources, lower than our limit of detection, resulting in a $\sim$50\% combined probability of non detection, 
or the anisotropic nature of the pulsar emission, with a probability of $\sim$19\% that the beam of all the four sources does not 
intersect our line of sight.


\begin{acknowledgements}
This work is supported by the Italian Space Agency, ASI-INAF I/088/06/0 contract for High Energy Astrophysics, and
                       by the RAS (Regione Autonoma della Sardegna), PO-FSE 2007-13, L.R. 7/2007.
\end{acknowledgements}


\onecolumn
\newpage
\onltab{3}{
\begin{table}[tt]
\caption{Parameters for the source, the observation and data analyses for \otto, \sette\ and \sax.}
\begin{center}
\begin{tabular}{lcccccc}
\hline
\hline
Pulsar names  & \multicolumn{2}{c}{\otto} &\multicolumn{2}{c}{\sette}&\multicolumn{2}{c}{\sax}\\
\hline
\hline
\multicolumn{2}{c}{Source parameters}  & &&\\
\hline
Right ascension (J2000.0) & \multicolumn{2}{c}{18$^{\rm h}$13$^{\rm m}$39\fsecs04} &\multicolumn{2}{c}{17$^{\rm h}$51$^{\rm m}$13\fsecs49} & \multicolumn{2}{c}{18$^{\rm h}$08$^{\rm m}$27\fsecs54}\\  
Declination (J2000.0) & \multicolumn{2}{c}{$-$33$^{\circ}$46$^{\prime}$22\farcs3}&\multicolumn{2}{c}{$-$30$^{\circ}$37$^{\prime}$23\farcs4} & \multicolumn{2}{c}{$-$36$^{\circ}$58$^{\prime}$44\farcs3}\\
Orbital period, $P_{\rm orb}$ (s) & \multicolumn{2}{c}{15388.7229(2)} &\multicolumn{2}{c}{2545.342(2)}&\multicolumn{2}{c}{7249.156980(4)}\\
Projected semi-major axis, $a\sin i$ (lt-ms) & \multicolumn{2}{c}{390.633(9)}  &\multicolumn{2}{c}{10.125(5)}&\multicolumn{2}{c}{62.812(2)}\\
Eccentricity, $e$ & \multicolumn{2}{c}{$< 2.4 \times 10^{-5}$}  &\multicolumn{2}{c}{$< 1.3 \times 10^{-3}$}&\multicolumn{2}{c}{$< 1.2 \times 10^{-4}$}\\
Spin period, $P_{\rm S}$ (s) & \multicolumn{2}{c}{0.0031811056698(1)} &\multicolumn{2}{c}{0.002297172972(2)}&\multicolumn{2}{c}{0.00249391975978(6)}\\
Mean spin period  derivative, $\dot{P}_{\rm S}$ (s s$^{-1}$) & \multicolumn{2}{c}{6.7(7) $\times \ 10^{-19}$}&\multicolumn{2}{c}{$-$1.95(10) $\times \ 10^{-18}$}&\multicolumn{2}{c}{4.7(9) $\times \ 10^{-19}$}\\
Ascending node passage, $T_0$ (MJD) & \multicolumn{2}{c}{52797.8101689(9)}&\multicolumn{2}{c}{52368.0129023(4)}&\multicolumn{2}{c}{52499.9602472(9)}\\
Distance, $d$ (kpc) & \multicolumn{2}{c}{8} & \multicolumn{2}{c}{6.7}&\multicolumn{2}{c}{3.5}\\
Nominal DM (pc cm$^3$) &  \multicolumn{2}{c}{$\sim$200}&\multicolumn{2}{c}{$\sim$400} & \multicolumn{2}{c}{$\sim$100}\\
References  & \multicolumn{2}{c}{1} & \multicolumn{2}{c}{2} & \multicolumn{2}{c}{3,4} \\
\hline
\multicolumn{2}{c}{Observation parameters}&&1$^{\rm st}$ &2$^{\rm nd}$&&\\
\hline
Central radio frequency, $\nu_{\rm obs}$ (GHz) &6.4105&8.4535&8.4535&8.4535&4.7495&6.3515\\
Data series time, $\Delta t$ (s)&25053&10669&17145&22864 &35685&35051\\
Sampling time, $\delta t_{\rm sam}$ ($\mu$s) &92&50&60&100&125&125\\
Samples number, $N_{\rm s}$&2$^{28}$&2$^{27}$&2$^{28}$&2$^{27}$&2$^{28}$&2$^{28}$\\
Nominal gain\tablefootmark{a}, $G$ (K Jy$^{-1}$)&0.46&0.59&\multicolumn{2}{c}{0.59}&\multicolumn{2}{c}{0.46}\\
System temperature\tablefootmark{a}, $T_{\rm sys}$ (K) &50&25&\multicolumn{2}{c}{25}&\multicolumn{2}{c}{50}\\
Number of 3 MHz channels, $N_{\rm c}$ &\multicolumn{2}{c}{192}&\multicolumn{2}{c}{192}&\multicolumn{2}{c}{192}\\
\hline
\multicolumn{2}{c}{Data analysis parameters}&&&&&\\
\hline
Number of DMs&186&151&195&127&135&57\\
DM range (pc cm$^3$)&\multicolumn{2}{c}{$\sim$ 50$\div$1000}&\multicolumn{2}{c}{$\sim$ 100$\div$1700}&\multicolumn{2}{c}{$\sim$ 20$\div$400}\\
Local DM (pc cm$^3$)&\multicolumn{2}{c}{$\sim$20}&\multicolumn{2}{c}{$\sim$260}&\multicolumn{2}{c}{$\sim$230}\\
Number of $P_{\rm orb}$ steps&\multicolumn{2}{c}{1}&\multicolumn{2}{c}{19}&\multicolumn{2}{c}{1}\\
Number of folding steps&\multicolumn{2}{c}{1}&\multicolumn{2}{c}{11}&\multicolumn{2}{c}{1}\\
\hline
\multicolumn{2}{c}{Other parameters}&&&&&\\
\hline
Sky temperature, $T_{\rm sky}$ (K)&0.06&0.03&\multicolumn{2}{c}{0.09}&0.13&0.06\\
Dispersion broadening, $\delta t_{\rm DM}$ ($10^{-8}$s)&9.5 $\times$ DM & 4.2 $\times$ DM&\multicolumn{2}{c}{4.2 $\times$ DM}&23 $\times$ DM&9.7 $\times$ DM\\
Scattering broadening, $\delta t_{\rm scatt}$ (s)&$9.3 \times 10^{-9}$ & $3.2 \times 10^{-9}$&\multicolumn{2}{c}{$4.4\times 10^{-7}$} &9.6 $\times 10^{-9}$&3 $\times 10^{-9}$\\
\hline
\hline
\end{tabular}
\tablefoot{\tablefoottext{a}{Values derived from the Parkes website: \protect\url{http://www.parkes.atnf.csiro.au/observing/documentation/}.}}
\tablebib{(1) \citet{pdb+07}; (2) \citet{pmb+08}; (3) \citet{hpc+09}; (4) \citet{bdm+06}.}
\end{center}
\label{tab:parameters}
\end{table}
}

\end{document}